\documentclass{article}

\usepackage{arxiv}

\usepackage[utf8]{inputenc} 
\usepackage[T1]{fontenc}    
\usepackage[hidelinks]{hyperref}       
\usepackage{url}            
\usepackage{booktabs}       
\usepackage{amsfonts}       
\usepackage{nicefrac}       
\usepackage{microtype}      
\usepackage{lipsum}		
\usepackage{graphicx}
\usepackage[numbers]{natbib}
\usepackage{doi}
\usepackage{multirow, longtable}

\title{Family-wise error rate control in clinical trials with overlapping populations}

\author{{Remi Luschei} \\
	Competence Center for Clinical Trials Bremen \\
	Institute for Statistics\\
	University of Bremen\\
	\texttt{rluschei@uni-bremen.de} \\
	\And
	{Werner Brannath} \\
	Competence Center for Clinical Trials Bremen\\
	Institute for Statistics \\
	University of Bremen\\
	\texttt{brannath@uni-bremen.de} \\
}

\renewcommand{\headeright}{}
\renewcommand{\undertitle}{}
\hypersetup{
pdftitle={Family-wise error rate control in clinical trials with overlapping populations},
pdfauthor={Remi Luschei, Werner Brannath},
}

\usepackage{amsmath, amsfonts, mathtools, csquotes, url, bbm, tikz, bm, amsthm, subcaption}
\usepackage[titletoc,toc,page]{appendix}
\usepackage{tikz}
\newcommand{\Pop}{\mathcal{P}}
\renewcommand{\S}{\mathcal{S}}
\newcommand{\R}{\mathbb{R}}

\newcommand{\E}{\mathbb{E}}
\newcommand{\I}{\mathcal{I}}
\newcommand{\Var}{\text{Var}}

\renewcommand{\P}{\mathbb{P}}

\newcommand{\FWER}{\text{FWER}}

\theoremstyle{definition}
\newtheorem{example}{Example}
\newcommand{\btheta}{\bm{\theta}}

\usepackage{setspace}
\begin{document}
\maketitle 

\begin{abstract}
We consider clinical trials with multiple, overlapping patient populations, that test multiple treatment policies specifically tailored to these populations. Such designs may lead to multiplicity issues, as false statements will affect several populations. For type I error control, often the family-wise error rate (FWER) is controlled, which is the probability to reject at least one true null hypothesis. If the joint distribution of the test statistics is known, the FWER level can be exhausted by determining critical values or adjusted $\alpha$-levels. The adjustment is typically done under the common ANOVA assumptions. However, the performed tests are then only valid under the rather strong assumption of homogeneous null effects, i.e., when the null hypothesis applies to all subpopulations and their intersections. We show that under cancelling null effects, when heterogeneous effects cancel out in some or all subpopulations, this procedure does not provide FWER control. We also suggest different alternatives and compare them in terms of FWER control and their power.
\end{abstract}


\section{Introduction}

\begin{figure}[h]
    \centering
    \begin{subfigure}{0.32\textwidth}
        \centering
        \begin{tikzpicture}
            \draw[->] (0,0) -- (0,2.25) node[midway, rotate=90, yshift=0.4cm] {Treatment effect};
            \draw[->] (0,0) -- (3,0) node[below] {};
            
            \draw[dashed] (0,1) -- (3,1);
            
            \fill[black!60] (1,2) circle (5pt); 
            \fill[black!20] (2,2) circle (5pt); 
            
            \node[below] at (1,0) {$\S_1$};
            \node[below] at (2,0) {$\S_2$};
            \node[right] at (0,1.5) {\tiny E $>$ C};
            \node[right] at (0,0.5) {\tiny E $<$ C};
        \end{tikzpicture}
        \caption{No SEH}
    \end{subfigure}
    \hfill
    \begin{subfigure}{0.32\textwidth}
        \centering
        \begin{tikzpicture}
            \draw[->] (0,0) -- (0,2.25) node[midway, rotate=90, yshift=0.4cm] {Treatment effect};
            \draw[->] (0,0) -- (3,0) node[below] {};
            
            \draw[dashed] (0,1) -- (3,1);
            
            \fill[black!60] (1,2) circle (5pt);
            \fill[black!20] (2,1.5) circle (5pt);
            
            \node[below] at (1,0) {$\S_1$};
            \node[below] at (2,0) {$\S_2$};
            \node[right] at (0,1.5) {\tiny E $>$ C};
            \node[right] at (0,0.5) {\tiny E $<$ C};
        \end{tikzpicture}
        \caption{Quantitative SEH}
    \end{subfigure}
    \hfill
    \begin{subfigure}{0.32\textwidth}
        \centering
        \begin{tikzpicture}
            \draw[->] (0,0) -- (0,2.25) node[midway, rotate=90, yshift=0.4cm] {Treatment effect};
            \draw[->] (0,0) -- (3,0) node[below] {};
            
            \draw[dashed] (0,1) -- (3,1);
            
            \fill[black!60] (1,2) circle (5pt);
            \fill[black!20] (2,0.4) circle (5pt);
            
            \node[below] at (1,0) {$\S_1$};
            \node[below] at (2,0) {$\S_2$};
            \node[right] at (0,1.5) {\tiny E $>$ C};
            \node[right] at (0,0.5) {\tiny E $<$ C};
        \end{tikzpicture}
        \caption{Qualitative SEH}
    \end{subfigure}    
    \caption{Different kinds of subgroup effect heterogeneity (SEH) between two subgroups $\S_1$ and $\S_2$, when comparing an investigational treatment $E$ to a control $C$ (figure adapted from \citet{wang})}
    \label{fig1}
\end{figure}
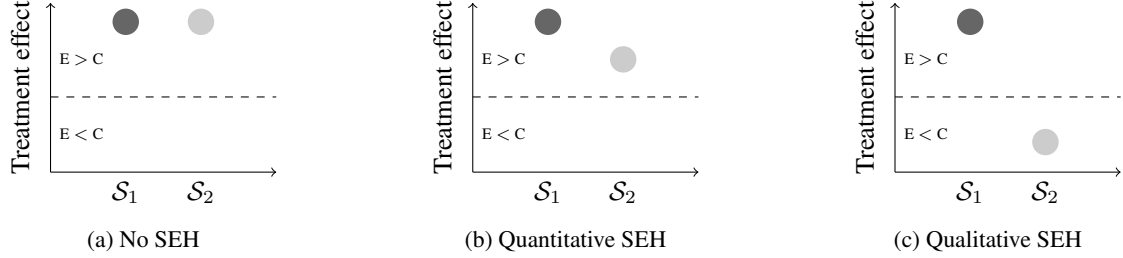

Clinical trials with multiple target populations have become increasingly important in recent years in the context of personalized medicine, which aims to find tailored treatments for individual patients. This introduces greater complexity in trial designs, as treatment effects may vary among patient subgroups. Subgroup effect heterogeneity can be either quantitative, when the subgroup effects are in the same direction but differ in magnitude, as in Figure \ref{fig1} (b), or qualitative, when the subgroup effects have opposite directions, being beneficial for some groups and harmful for others, see Figure \ref{fig1} (c) (\cite{wang, gabler}). In practice, the target populations of a trial may include multiple subgroups and be overlapping, meaning that individual patients can belong to multiple populations simultaneously. Examples include enrichment trials, umbrella and basket trials, as well as platform trials (see e.g.\ \citet{antognini}). This leads to a multiplicity issue, as patiens may be subjected to several treatment decisions, potentially resulting in an exposure to inefficient therapies. 

To address this, \citet{sun} recommend controlling the family-wise error rate (FWER), ensuring that the probability of making one or more false discoveries across all tested hypotheses remains bounded. Alternatively, one could also control the population-wise error rate (PWER) introduced by \citet{brannath}, which is an average of FWERs that are restricted to the subpopulations, and thereby becomes more liberal than the FWER. In the following section, we briefly describe how to control the FWER in a setting with overlapping populations. A similar method is applicable for the PWER.

\subsection{FWER control in a single stage design} \label{sec:fwer}

Let us consider a trial with a patient population $\Pop$ that can be partitioned into disjoint subpopulations $\S_1, \dots, \S_n$. We are interested in testing treatments within certain combinations of these subpopulations, represented by a finite collection of index sets $\I \subseteq 2^{\{1, \dots, n\}}$. In each $\Pop_I \coloneqq \bigcup_{i \in I} \S_i$, $I \in \I$, we want to test an experimental treatment $E_I$ in comparison to a control treatment $C$. Let $\mu_{\S_i, T} \in \R$ denote the expected response to treatment $T \in \mathcal{T}_i$ within subpopulation $\S_i$, where $\mathcal{T}_i = \{E_I : i \in I\} \cup \{C\}$ is the set of treatments administered in $\S_i$. We assume throughout that higher response values correspond to better outcomes. The treatment effect of $E_I$ in $\Pop_I$ is defined as
\begin{align} \label{eq:contr}
\theta_I = \mu_{\Pop_I, E_I} - \mu_{\Pop_I, C} = \sum_{i \in I}\frac{ \pi_{\S_i}}{\pi_{\Pop_I}} \theta_{\S_i, E_I},
\end{align}
where $\theta_{\S_i, E_I} = \mu_{\S_i, E_I} - \mu_{\S_i, C}$ is the mean treatment effect of $E_I$ in $\S_i$, and $\pi_{\S_i}$ denotes the prevalence of $\S_i$ in $\Pop$ (so that $\pi_{\Pop_I} = \sum_{i \in I} \pi_{\S_i}$). Our null hypotheses of interest are:
\[
H_I: \theta_I \leq 0, \quad \text{for all } I \in \I.
\]

The FWER is defined as the probability to reject at least one true $H_I$ for $I \in \I$. It is said to be strongly controlled at level $\alpha \in (0, 1)$ if
\[
\FWER_{\btheta} = \P\left(\bigcup_{I \in \I_0(\btheta)} \{\text{Reject } H_I\} \right) \leq \alpha \quad \text{for all } \btheta = (\theta_I)_{I \in \I},
\]
where $\I_0(\btheta) \coloneqq \{I \in \I : \theta_I \leq 0\}$ is the set of true null hypotheses under the configuration $\btheta$. As observed by \citet{ondra}, overlapping populations $\Pop_I$ generally induce correlation among the test statistics, such that nonparametric procedures like the Bonferroni correction may be overly conservative. If the joint distribution of the test statistics is known, more powerful procedures can be constructed. Suppose that each hypothesis $H_I$ is tested with a test statistic $Z_I$ and a common rejection threshold $c$. Then the FWER under a specific configuration $\btheta$ equals
\[
\FWER_{\btheta}(c) = \P\left( \bigcup_{I \in \I_0(\btheta)} \{Z_I > c\} \right).
\]
The maximal FWER typically occurs under the global null hypothesis $\btheta = \boldsymbol{0}$, i.e., when $\theta_I = 0$ for all $I \in \I$, as is the case for many common test statistics such as contrast $z$- or $t$-statistics under normality assumptions (see, e.g., Theorem 1 in \cite{luschei}). In this case, strong and exhaustive control of the FWER at level $\alpha \in (0,1)$ can be achieved by selecting the threshold $c$ that satisfies $\FWER_{\boldsymbol{0}}(c) = \alpha$. Alternatively, one can compute FWER-adjusted p-values by $p_I = \FWER_{\boldsymbol{0}}(z_I^{\text{obs}})$, where $z_I^{\text{obs}}$ denotes the observed value of $Z_I$.

\subsection{Problem formulation}

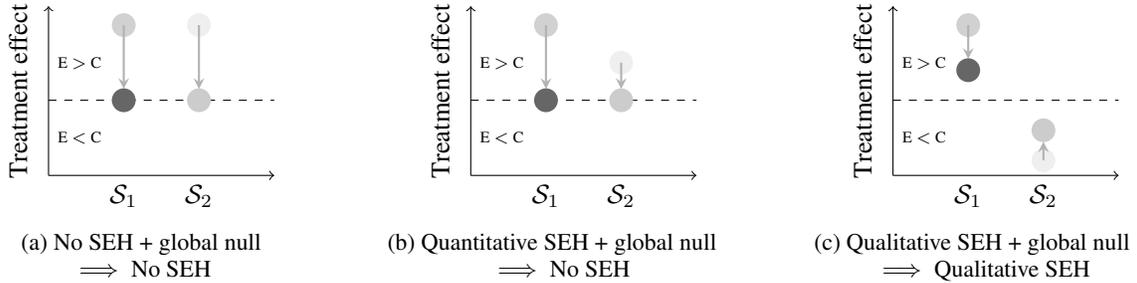
\begin{figure}[h]
    \centering
    \begin{subfigure}{0.32\textwidth}
        \centering
        \begin{tikzpicture}
            \draw[->] (0,0) -- (0,2.25) node[midway, rotate=90, yshift=0.4cm] {Treatment effect};
            \draw[->] (0,0) -- (3,0) node[below] {};
            
            \draw[dashed] (0,1) -- (3,1);
            
            \fill[black!60, opacity=0.3] (1,2) circle (4.5pt);
            \fill[black!20, opacity=0.3] (2,2) circle (4.5pt);
            
            \draw[->, gray!60, thick, >=stealth] (1,2) -- (1,1.15);
            \draw[->, gray!60, thick, >=stealth] (2,2) -- (2,1.15);            
            
            \fill[black!60] (1,1) circle (4.5pt);
            \fill[black!20] (2,1) circle (4.5pt);
            
            \node[below] at (1,0) {$\S_1$};
            \node[below] at (2,0) {$\S_2$};
            \node[right] at (0,1.5) {\tiny E $>$ C};
            \node[right] at (0,0.5) {\tiny E $<$ C};
        \end{tikzpicture}
        \caption{No SEH + global null \\\centering $\implies$ No SEH}
    \end{subfigure}
    \hfill
    \begin{subfigure}{0.32\textwidth}
        \centering
        \begin{tikzpicture}
            \draw[->] (0,0) -- (0,2.25) node[midway, rotate=90, yshift=0.4cm] {Treatment effect};
            \draw[->] (0,0) -- (3,0) node[below] {};
            
            \draw[dashed] (0,1) -- (3,1);
            
            \fill[black!60, opacity=0.3] (1,2) circle (4.5pt);
            \fill[black!20, opacity=0.3] (2,1.5) circle (4.5pt);
            
            \draw[->, gray!60, thick, >=stealth] (1,2) -- (1,1.15);
            \draw[->, gray!60, thick, >=stealth] (2,1.5) -- (2,1.15);
            
            \fill[black!60] (1,1) circle (4.5pt);
            \fill[black!20] (2,1) circle (4.5pt);
            
            \node[below] at (1,0) {$\S_1$};
            \node[below] at (2,0) {$\S_2$};
            \node[right] at (0,1.5) {\tiny E $>$ C};
            \node[right] at (0,0.5) {\tiny E $<$ C};
        \end{tikzpicture}
        \caption{Quantitative SEH + global null \\\centering $\implies$ No SEH}
    \end{subfigure}
    \hfill
    \begin{subfigure}{0.32\textwidth}
        \centering
        \begin{tikzpicture}
            \draw[->] (0,0) -- (0,2.25) node[midway, rotate=90, yshift=0.4cm] {Treatment effect};
            \draw[->] (0,0) -- (3,0) node[below] {};
            
            \draw[dashed] (0,1) -- (3,1);
            
            \fill[black!60, opacity=0.3] (1,2) circle (4.5pt);
            \fill[black!20, opacity=0.3] (2,0.2) circle (4.5pt);
            
            \draw[->, gray!60, thick, >=stealth] (1,2) -- (1,1.55);
            \draw[->, gray!60, thick, >=stealth] (2,0.2) -- (2,0.45);
            
            \fill[black!60] (1,1.4) circle (4.5pt);
            \fill[black!20] (2,0.6) circle (4.5pt);
            
            \node[below] at (1,0) {$\S_1$};
            \node[below] at (2,0) {$\S_2$};
            \node[right] at (0,1.5) {\tiny E $>$ C};
            \node[right] at (0,0.5) {\tiny E $<$ C};
        \end{tikzpicture}
        \caption{Qualitative SEH + global null \\\centering $\implies$ Qualitative SEH}
    \end{subfigure}
     \caption{Subgroup effect heterogeneity under the global null hypothesis. The arrows show the adjustment of the effects from Figure \ref{fig1} so that the treatment effect in the overall population $\S_1 \cup \S_2$ equals 0. The prevalences of $\S_1$ and $\S_2$ are assumed to be equal.
}
     \label{fig2}
\end{figure}

As noted by \citet{ondra}, many authors assume normally distributed patient outcomes and apply the procedure described above to the contrasts defined in (\ref{eq:contr}) under the ANOVA assumptions. We will demonstrate that under a qualitative effect heterogeneity between the subgroups $\S_i$, this approach may fail to control the FWER (and similar error rates such as the PWER). This is due to the fact that a qualitative effect heterogeneity may remain under the global null hypothesis due to cancelling subgroup effects (see the illustration in Figure \ref{fig2} (c)). In such a situation, the overall treatment effect can be zero in a population $\Pop_I$, but the subgroup-specific effects are nonzero and heterogeneous in sign. 

As a result, the contrasts $\theta_I$ cannot be reliably estimated from the data under the null hypothesis in the ANOVA framework. The reason is that they depend on the unknown subpopulation prevalences, while the ANOVA model estimates the contrasts conditional on the observed subgroup sample sizes. This leads the ANOVA to implicitly reweight the subgroup effects in a way that does not reflect the true population mixture. While that is negligible under homogeneous effects (which are uniformly zero under the global null; see Figure \ref{fig2} (b)), under heterogeneity it distorts the expected value of the contrasts from zero and thereby creates a bias which questions error control, as will be detailed in Section \ref{sec:anova}. 

\subsection{Overview of the paper}

We present the ANOVA model in Section \ref{sec:anova} and show an example where FWER control is missed. In Section \ref{sec:tests}, we propose different alternative methods to adress the issue described above and in Section \ref{sec:sim}, we compare them with respect to FWER control and power in a simulation study. In Section \ref{sec:ci} we construct corresponding confidence intervals if applicable. In Section \ref{sec:ex} we apply the proposed tests to a real data example. The note ends with a discussion in Section \ref{sec:dis}.

\section{The ANOVA subpopulation model} \label{sec:anova}

For each treatment $T \in \mathcal{T}_i$, let $n_{\mathcal{S}_i, T}$ denote the number of patients from subgroup $\mathcal{S}_i$ assigned to treatment $T$. Similarly, let $ n_{\Pop_I, T}$ represent the number of patients in population \( \Pop_I \) receiving treatment $T \in \{E_I, C\}$. Let $N$ denote the total sample size. In the ANOVA model, the observations in every $\S_i$ are assumed to be of the form \[Y_{\S_i, T}^{(k)} = \mu_{\S_i, T} + \varepsilon^{(k)}, \quad k=1, \dots, n_{\S_i, T}, \quad T \in \mathcal{T}_i,\]  where $\varepsilon^{(k)} \sim N(0, \sigma^2)$ is the residual of patient $k$, and the residuals are assumed to be stochastically independent across patients. The test statistic for $H_I$ is
\begin{align} \label{eq:t-anova}
Z_I = \frac{\bar{Y}_{\Pop_I, E_I}-\bar{Y}_{\Pop_I, C}}{\sigma\sqrt{V_I}} \quad \text{with} \quad V_I = n_{\Pop_I,E_I}^{-1} + n_{\Pop_I,C}^{-1},
\end{align}
where $\bar{Y}_{\Pop_I, E_I}$ and $\bar{Y}_{\Pop_I, C}$ are the average responses under treatments $E_I$ and $C$ in the population $\Pop_I$. For simplicity we assume that the  variance $\sigma^2$ is known. Conditional on the subgroup sample sizes, the test statistics follow a multivariate normal distribution with the location parameter $\boldsymbol{\nu} = (\nu_I)_{I \in \I}$ given by 
\begin{align}
\nu_I &= \frac{1}{\sigma\sqrt{V_I}}\left(\sum_{i \in I} \frac{n_{\S_i,E_I}}{n_{\Pop_I, E_I}}\mu_{\S_i, E_I} - \frac{n_{\S_i,C}}{n_{\Pop_I, C}}\mu_{\S_i, C} \right), \label{eq1} 
\end{align}
and the correlation matrix $\boldsymbol{\Sigma} = (\Sigma_{IJ})$ given by 
\begin{align}
\Sigma_{IJ} &= \frac{1}{\sqrt{V_IV_J}}\sum_{i \in I \cap J} \frac{n_{\S_i,E_I}}{n_{\Pop_I, E_I}n_{\Pop_J,E_I}} \mathbbm{1}(E_I = E_J) +\frac{n_{\S_i,C}}{n_{\Pop_I, C}n_{\Pop_J,C}}, \quad I \neq J \label{eq2}
\end{align} 
where $\mathbbm{1}(E_I = E_J)$ is the indicator function that equals 1 if $E_I = E_J$ and 0 otherwise. If $\sigma^2$ is unknown and estimated as a pooled variance, the test statistics follow a multivariate $t$-distribution with parameters $\boldsymbol{\nu}, \boldsymbol{\Sigma}$ and $N-s$ degrees of freedom, where $s$ is the total number of combinations of subpopulations and treatments. 

As we have seen in Section \ref{sec:fwer}, strong and exhaustive control of the FWER requires knowledge of the parameters $\boldsymbol{\nu}$ and $\boldsymbol{\Sigma}$ under the global null hypothesis $\boldsymbol{\theta} = \boldsymbol{0}$. The covariance matrix $\boldsymbol{\Sigma}$ is known by design, as it depends only on the (known) subgroup sample sizes. In contrast, $\boldsymbol{\nu}$ cannot be exactly determined because it depends on the unknown constants $\mu_{\S_i, E_I}$ and $\mu_{\S_i, C}$. However, when excluding the possibility that the subpopulation effects $\theta_{\S_i} = \mu_{\S_i, E_I}-\mu_{\S_i, C}$  have opposite signs,  $\boldsymbol{\nu}$ converges to $\boldsymbol{0}$ under the null hypotheses as the total sample size $N$ approaches infinity.  This follows from the fact that the sample proportions $n_{\S_i,T}/n_{\Pop_I,T}$ converge to their population counterparts $\pi_{\S_i}/\pi_{\Pop_I}$. Without this assumption, it is possible for subpopulation effects $\theta_{\S_i}$ to be opposite and to cancel each other out when weighted by their prevalences. In such cases, the overall null hypothesis $\boldsymbol{\theta} = \boldsymbol{0}$ may still hold when $\boldsymbol{\nu}$ does not necessarily converge to zero. Consequently, we would be unable to approximate $\boldsymbol{\nu}$, even for the purposes of an asymptotic test. We demonstrate this in the following example.

\begin{example} \label{ex1}

We consider a simple example with three subpopulations $\S_1, \S_2, \S_3$, and two target populations defined as follows: $\Pop_1 = \S_1 \cup S_2 \cup \S_3$ and $\Pop_2 = S_2 \cup \S_3$. In both $\Pop_1$ and $\Pop_2$, the same treatment $E$ is tested against the control $C$. For simplicity, we assume a balanced allocation of the patients in the subpopulations, i.e.\ $n_{\S_i, T} = n_{\S_i, C}$ for every $i=1,2,3$, which can be achieved approximately with a stratified randomization. Additionally, we assume a common residual variance of $\sigma^2=1/4$. From formula (\ref{eq1}) we then get that
\begin{align} \label{eq3}
\nu_1 =  \frac{n_{\S_1}\theta_{\S_1}+n_{\S_2}\theta_{\S_2}+n_{\S_3}\theta_{\S_3}}{\sqrt{N}} \quad \text{and} \quad
\nu_2 = \frac{n_{\S_2}\theta_{\S_2}+n_{\S_3}\theta_{\S_3}}{\sqrt{n_{\S_2}+n_{\S_3}}}.
\end{align}
Suppose the subpopulation prevalences are given by $\pi_{\S_1} = \pi_{\S_2}= 1/6$, $\pi_{\S_3} = 2/3$ and the true treatment effects are $\theta_{\S_1}=0$, $\theta_{\S_2}=4$, $\theta_{\S_3}=-1$. In this case, we find that $\boldsymbol{\theta}=\boldsymbol{0}$. By the central limit theorem, the vector $\boldsymbol{\nu}$ converges in distribution as follows (see Appendix \ref{app:ex1} for the details):
\begin{align*}
\nu_1 &= \frac{4n_{\S_2} - n_{\S_3}}{\sqrt{N}} \xrightarrow[N \to \infty]{d} N(0, 10/3), \quad \nu_2 = \frac{4n_{\S_2} - n_{\S_3}}{\sqrt{5N/6}} \cdot \sqrt{\frac{5N/6}{n_{\S_2} + n_{\S_3}}} \xrightarrow[N \to \infty]{d} N(0, 4).
\end{align*}
To investigate the effect of these fluctuations of $\boldsymbol{\nu}$ on FWER control, we conduct a simulation study where in each iteration, we generate random sample sizes $n_{\S_i}$ from the multinomial distribution with parameters $N$ and $\boldsymbol{\pi}$. We then calculate the rejection boundary $c$, assuming that $\boldsymbol{\nu}=\boldsymbol{0}$. As significance level we take $\alpha = 0.025$. We then calculate the resulting true FWER using the true $\boldsymbol{\nu}$ from equation (\ref{eq3}). We repeat this procedure $10^4$ times. On average, we observe that the true FWER increases to values between 0.18 and 0.19, depending on the total sample size which we vary between 250 and 1000 patients, which are realistic sample sizes for multi-population studies. A more systematical analysis will be provided in Section \ref{sec:sim}.
\end{example}

\section{Test procedures} \label{sec:tests}

\subsection{Bootstrap approximation} \label{sec:boot}

We propose a parametric bootstrap procedure to approximate the distribution of the test statistics given in (\ref{eq:t-anova}) under the global null hypothesis. Let $[n] \coloneqq \{1, \dots, n\}$. In each iteration, the subpopulation sample sizes, denoted by $n_{\S_i}^*$, are redrawn from the multinomial distribution with number of trials $N$ and probabilities $(n_{\S_i}/N)_{i \in [n]}$, corresponding to the observed strata proportions. For every $i \in [n]$ and $T \in \mathcal{T}_i$, the treatment-wise allocation numbers are then set as $n_{\S_i, T}^* = n_{\S_i}^* \delta_{\S_i, T}$, where $\delta_{\S_i, T} = n_{\S_i, T}/n_{\S_i}$ denotes the observed allocation rate in the original sample. Next, the subpopulation means are resampled from independent normal distributions with parameters $\mu_{S_i, T}^0$ and $\sigma_{S_i, T}^{*2}$ that are chosen as follows: 
\begin{itemize}
\item $\boldsymbol{\mu}^0 = (\mu^0_{S_i, T})_{i \in [n],\, T \in \mathcal{T}_i}$ is defined as the orthogonal projection of the observed subpopulation means $\boldsymbol{\bar Y} = (\bar Y_{S_i, T})_{i \in [n],\, T \in \mathcal{T}_i}$ onto the linear subspace
\[
\mathcal{M}_0 = \{\boldsymbol{\mu} : \boldsymbol{r}_I^\top \boldsymbol{\theta}_{\mathcal{S}, I} = 0 \text{ for all } I \in \mathcal{I} \},
\]
with $\boldsymbol{r}_I = (n_{S_i}/n_{\Pop_I})_{i \in I}$ and $\boldsymbol{\theta}_{\mathcal{S}, I} = (\mu_{S_i, E_I} - \mu_{S_i, C})_{i \in I}$.
$\mathcal{M}_0$ contains all $\boldsymbol{\mu}$ that satisfy the constraints of the (estimated) global null hypothesis under the LFC. We find $\boldsymbol{\mu}^0$  e.g.\ as the residuals from a linear regression of $\boldsymbol{\bar Y}$ with one covariate vector for each $I \in \mathcal{I}$, where the covariate for $I$ assigns the values $(\boldsymbol{r}_I)_i$ to the coordinates corresponding to $(S_i, E_I)$, $i \in I$, and $-(\boldsymbol{r}_I)_i$ at the coordinates corresponding to $(S_i, C)$, with zeros elsewhere. 

\item We define $\sigma^{*2}_{S_i, T} \coloneqq \sigma^2/n_{\S_i,T}^*$. Hereby, the variance $\sigma^2$ is estimated as a pooled variance from the original sample if necessary.
\end{itemize}
This gives us a set of bootstrapped test statistics $Z_I^*$. An FWER-adjusted $p$-value for $H_I$ can then be calculated as \[
p_I = \frac{\#\left\{ \max_{I' \in \I}(Z_{I'}^*) \geq z_I^\text{obs} \right\}}{n_\text{boot}},
\] where $z_I^\text{obs}$ is the originally observed test statistic, and $n_\text{boot}$ is the number of bootstrap iterations.

We note that the test statistics in (\ref{eq:t-anova}) fulfill the conditions of the \enquote{smooth function model} introduced by \citet{hall}. It follows that the above bootstrap tests are second order accurate, meaning that the approximation error has order $O(N^{-1})$, and that the FWER is asymptotically controlled.

\subsection{Marginal tests} \label{sec:marg}

To account for a potential quantitative subgroup effect hetergeneity in the test statistics, in every population $\Pop_I$ we consider observations of the form 
\begin{equation}\label{eq:marg_obs}
\begin{split}
Y_{\Pop_I, T}^{(k)} &= \mu_{\Pop_I, T}+( \mu_{\S_{i^*}, T} - \mu_{\Pop_I, T}) + \varepsilon^{(k)}, \quad k=1, \dots, n_{\Pop_I, T}, \quad T \in \{E_I, C\}\\
&= \mu_{\Pop_I, T}+ \gamma_{\S_{i^*}, T} + \varepsilon^{(k)},
\end{split}
\end{equation}
where the index $i^*$ is now seen as a random variable which takes the values $i \in I$ with probabilities $\pi_{\S_i}/\pi_{\Pop_I}$. Thus, a discrete random variable $\gamma_{\S_{i^*}, T}$ with a positive variance is added to the expected response in population $\Pop_I$. For the residuals we still assume $\varepsilon^{(k)} \sim N(0, \sigma^2)$, and independence from the $\gamma_{\S_{i^*}, T}$. The test statistics are then \[Z_I= \frac{\bar{Y}_{\Pop_I, E_I}-\bar{Y}_{\Pop_I, C}}{\sqrt{\sigma_{\Pop_I, E_I}^2/n_{\Pop_I,E_I}+ \sigma_{\Pop_I,C}^2/n_{\Pop_I,C}}}, \quad I \in \I\] where $\sigma_{\Pop_I, T}^2 = \Var(\gamma_{\S_{i^*}, T}) + \sigma^2$ is the variance in population $\Pop_I$ under treatment $T \in \{E_I, C\}$. The expected value of $Z_I$ is now $\nu_I = \theta_I$, and the correlations $\Sigma_{IJ}$ are the expectation of formula (\ref{eq2}) over the subgroup sample sizes. We will therefore estimate the correlations as in (\ref{eq2}).  We approximate the distribution of $Z_I$ by a $t$-distribution with degrees of freedom according to \citet{satterthwaite}, \[df_I= \frac{(\sigma^2_{\Pop_I,E_I}/n_{\Pop_I,E_I}+\sigma^2_{\Pop_I,C}/n_{\Pop_I,C})^2}{(\sigma^2_{\Pop_I,E_I}/n_{\Pop_I,E_I})^2/\left(n_{\Pop_I,E_I}-1\right)+(\sigma^2_{\Pop_I,C}/n_{\Pop_I,C})^2/\left(n_{\Pop_I,C}-1\right)}.\] As shown by \citet{hasler}, FWER control can now be reached by computing individual critical values $c_I$ for each $Z_I$ from the $n$-variate $t$-distribution with $df_I$ degrees of freedom: 
\begin{align} \label{eq4}
1- \Phi_{\boldsymbol{\theta}, \boldsymbol{\Sigma}, df_I}(c_I, \dots, c_I)  = \alpha 
\end{align}
This especially means that the maximal FWER obtained under $\boldsymbol{\theta} = \boldsymbol{0}$ will be controlled at least approximately. 

The marginal tests can also be performed under the more realistic assumption of heterogeneous variances $\sigma^2_{S_i, T}$ across the subpopulations and treatments. The only difference is then that the correlations of the test statistics depend on these variances and must also be estimated: 
\begin{align*}
\Sigma_{IJ} &= \frac{1}{\sqrt{V_IV_J}}\sum_{i \in I \cap J} \frac{n_{\S_i,E_I}\sigma^2_{\S_i, E_I}}{n_{\Pop_I, E_I}n_{\Pop_J,E_I}} \mathbbm{1}(E_I = E_J) +\frac{n_{\S_i,C}\sigma^2_{\S_i, C}}{n_{\Pop_I, C}n_{\Pop_J,C}} \quad \text{with} \quad V_I = \sum_{i \in I} \frac{n_{\S_i, E_I}\sigma_{\S_i, E_I}^2}{n_{\Pop_I, E_I}^2}  + \frac{n_{\S_i, C}\sigma_{\S_i, C}^2}{n_{\Pop_I, C}^2} 
\end{align*} 

Alternatively, the distribution of the marginal test statistics could be approximated via the bootstrap procedure presented in Section \ref{sec:boot}. Asymptotic FWER-control would then also apply to the marginal tests for the same reasons.

\subsection{Shrinkage method} \label{sec:shrink}

The marginal tests from Section \ref{sec:marg} may possibly become too conservative when the subgroup effects are highly heterogeneous, due to overestimation of the population variances. A natural solution to this problem would be to apply a shrinkage method that reduces the estimated population variances. We can achieve this by shrinking the subgroup means $\mu_{\S_i, T}$, $T \in \{E_I, C\}$, towards their arithmetic mean, e.g.\ by taking their James–Stein estimator \[\hat{\mu}_{\S_i, T} = 1- \frac{\#I-2}{\sum_{j\in I} \bar{Y}_{\S_j, T}^2 n_{\S_j, T}/\sigma^2}(\bar{Y}_{\S_i, T}-\bar{\bar{Y}}_T) + \bar{\bar{Y}}_T,\] where $\bar{\bar{Y}}_T$ is the sample mean of the $\bar{Y}_{\S_i, T}$, $i \in I$ (see e.g.\ \citet{taketomi}). Then we can calculate a shrinked variance from \[\hat{\sigma}^2_{\Pop_I, T} = \widehat{\Var}(\gamma_{\S_{i^*}, T}) + \sigma^2 =  \sum_{i \in I} \frac{n_{\S_i, T}}{n_{\Pop_I}} \hat{\mu}_{\S_i, T}^2 - \left(\sum_{i \in I} \frac{n_{\S_i,T}}{n_{\Pop_I}} \hat{\mu}_{\S_i, T} \right)^2 + \sigma^2\] and plug it into the test statistics. Note that the shrinkage only works in populations with at least three strata. For $\#I = 2$, the method reduces to the ANOVA tests from section \ref{sec:anova}.

\subsection{Stratified effect estimate}

It may be more efficient to calculate the mean differences within the strata first, and then weight them according to the respective strata sizes. That is, to use 
\begin{align} \label{eq:strat_est} 
\hat{\theta}_I =\sum_{i \in I} \frac{n_{S_i}}{n_{\Pop_I}} \hat{\theta}_{\S_i, E_I} = \sum_{i \in I} \frac{n_{S_i}}{n_{\Pop_I}} \left(\bar{Y}_{\S_i, E_I} - \bar{Y}_{\S_i, C}\right),
\end{align}
as effect estimates. One can quickly see that $\hat{\theta}_I$ converges almost surely to the true $\theta_I$ for $N \to \infty$, due to the almost sure convergence of the fractions $n_{\S_i}/N$ to $\pi_{\S_i}$. Moreover, $\hat{\theta}_I$ corresponds to a double robust effect estimator in the causal inference framework, specifically the augmented inverse-propensity weighting estimator (IPWE), where the response model includes subpopulation membership as a covariate (see e.g.\ \citet{hernan}). By standardizing, we obtain as test statistics \[Z_I = \frac{\hat{\theta}_I}{n_{\Pop_I}^{-1}\sqrt{\sigma^2 \sum_{i \in I} (\delta_{\S_i, E_I}^{-1} +\delta_{\S_i, C}^{-1})n_{\S_i} +  \hat{\boldsymbol{\theta}}_{\S,I} \hat{\boldsymbol{\Sigma}} \hat{\boldsymbol{\theta}}_{\S,I}^T}},\] where $\delta_{\S_i, T}$ denotes the observed allocation rate to treatment $T \in \{E_I, C\}$ in subpopulation $\S_i$, and where $\hat{\boldsymbol{\theta}}_{\S,I} = (\hat{\theta}_{\S_i, E_I})_{i \in I}$ and $\hat{\boldsymbol{\Sigma}}_I$ is the covariance matrix of the multinomial distribution $M(n_{\Pop_I}, \boldsymbol{\pi}_I)$, for $\boldsymbol{\pi}_I=(\pi_{\S_i})_{i \in I}$, which is estimated from the observed sample sizes. 
The calculation of the variance of $\hat{\theta}_I$ can be found in Appendix \ref{app:strat}. 

We will apply the bootstrap procedure presented in section \ref{sec:boot} to determine the joint null distribution of the $Z_I$ for FWER control. We note that the estimate given in (\ref{eq:strat_est}) can be written as a smooth function of the strata-wise sample sizes and observations, such that it also fits the smooth function model and asymptotical error control is reached.

\subsection{Random effects model}

Another approach to account for heterogeneous subgroup effects is a random effects model, where the observations in every population $\Pop_I$ are modeled as
\[Y_{\Pop_I}^{(k)} = \mu_{\Pop_I, C} + A \theta_{\Pop_I} + \gamma_{\S_i, C}^{(k)} + A \gamma_{\S_i, E_I}^{(k)} + \varepsilon_{\Pop_I}^{(k)}, \quad  \varepsilon_{\Pop_I}^{(k)} \sim N(0, \sigma^2), \quad k=1, \dots, n_{\Pop_I}.\]
Here $\gamma_{\S_i, C}$ and $\gamma_{\S_i, E_I}$ are two random effects associated with the subpopulation $\S_i$, and $A = \mathbbm{1}(T = E_I)$  is the indicator for the treatment assignment (1 for $E_I$, 0 for control). The null hypotheses $H_I$ are tested using Wald tests on the fixed effects. As we do not know the joint distribution of the test statistics, we apply a Bonferroni correction to adjust for multiplicity.

\section{Simulation study} \label{sec:sim}

We conduct a systematic comparison of the FWER and the multiple power (i.e.\ the expected proportion of correct rejections) of the different tests presented in Section \ref{sec:tests} using simulation studies. All programs are written in R. The corresponding R script files are available at the following link: \url{https://github.com/rluschei/fwer-seh}

\subsection{General setup}

As in Example \ref{ex1}, we consider two distinct target populations, $\Pop_1$ and $\Pop_2$, constructed from a common set of three disjoint subpopulations. The target populations may be nested, for example, $\Pop_1 = \S_1 \cup \S_2 \cup \S_3$ and $\Pop_2 = \S_2 \cup \S_3$, or partially overlapping, such as $\Pop_1 = \S_1 \cup \S_2$ and $\Pop_2 = \S_2 \cup \S_3$. We assume that the same investigational treatment is tested in $\Pop_1$ and $\Pop_2$. In each simulation run, we begin by generating the prevalences of the three subpopulations. To do this, we draw three independent random numbers uniformly from the interval $[0,1]$ and normalize them so that they sum to one. Next, we assign treatment effects to the subgroups. For FWER investigation, we start by generating the treatment effect for a subgroup that lies in the overlap of $\Pop_1$ and $\Pop_2$. This effect is drawn uniformly from the interval $[-1, 1]$ and scaled by a pre-specified \emph{effect heterogeneity factor (EHF)}, such as 0, 1, or 10, to reflect varying degrees of treatment effect heterogeneity. The treatment effects for the remaining subgroups are then computed by solving the linear equation system:
\[
\theta_{\Pop_I} = \pi_{\Pop_I}^{-1} \sum_{i \in I} \pi_{\S_i} \theta_{\S_i} = 0, \quad I \in \I,
\] so that the global null hypothesis holds. It is uniquely solvable as the effect in the overlap of $\Pop_1$ and $\Pop_2$ has already been specified. To investigate power, however, treatment effects for all subgroups are independently drawn from the interval $[-\text{EHF}, \text{EHF}]$ under the constraint that they fall under the alternative hypotheses. To introduce some variability in the absolute response levels, we also define a \emph{control group heterogeneity factor (CHF)}, with values such as 0, 1, or 10. This factor determines the expected control group responses: one subgroup has an expected response of 0, another receives a response equal to the CHF, and a third subgroup receives twice that amount. The residual variance is fixed at $\sigma^2 = 0.25$ across all subgroup-treatment combinations. We repeat this procedure 100 times, thereby simulating 100 different studies. For each study, we simulate its FWER and its power with 1000 simulation runs as detailed in Sections \ref{sec:sim-fwer} and \ref{sec:sim-power}.

\subsection{Simulated FWER} \label{sec:sim-fwer}

To simulate the FWER of a study, the subgroup sample sizes are independently redrawn in a simulation loop from the multinomial distribution with total sample size $N \in \{250,500, 1000\}$ and prevalence vector $\boldsymbol{\pi}$. We then generate the normally distributed data in the subgroups, assuming an equal treatment allocation, and apply the different tests from Sections \ref{sec:anova} and \ref{sec:tests} to both $\Pop_1$ and $\Pop_2$. To approximate the FWER, we calculate the proportion of iterations in which at least one null hypothesis is rejected. We set the significance level for FWER control to $\alpha = 0.025$, since all tests are done one-sided, and do a total of 1000 bootstrap sample draws per simulation run (when applicable). The resulting mean FWER estimates, for $N=500$ patients and with two nested populations, are reported in Table \ref{tab:fwer}.

\begin{table}[ht]
\centering
\begin{tabular}{cc|ccccccc}
\toprule
 EHF & CHF & anova+t & anova+boot & marg+t & marg+boot & marg+shr+boot & strat+boot & rem  \\
\midrule
0  & 0   & 0.0255 & 0.0264 & 0.0249 & 0.0268 & 0.0264 & 0.0266 & 0.0098 \\
0  & 1   & 0.0255 & 0.0264 & 0.0031 & 0.0265 & 0.0264 & 0.0266 & 0.0043 \\
0  & 10  & 0.0255 & 0.0264 & 0.0000 & 0.0261 & 0.0264 & 0.0266 & 0.0043 \\
1  & 0   & 0.0650 & 0.0292 & 0.0158 & 0.0297 & 0.0291 & 0.0302 & 0.0006 \\
1  & 1   & 0.0650 & 0.0292 & 0.0051 & 0.0292 & 0.0292 & 0.0302 & 0.0005 \\
1  & 10  & 0.0650 & 0.0292 & 0.0000 & 0.0293 & 0.0292 & 0.0302 & 0.0005 \\
10 & 0   & 0.3145 & 0.0337 & 0.0033 & 0.0386 & 0.0337 & 0.0393 & 0.0001 \\
10 & 1   & 0.3145 & 0.0337 & 0.0030 & 0.0400 & 0.0337 & 0.0393 & 0.0000 \\
10 & 10  & 0.3145 & 0.0337 & 0.0018 & 0.0400 & 0.0337 & 0.0393 & 0.0000 \\
\bottomrule
\end{tabular}
\vspace{10pt}
\caption{Simulated FWER for different effect heterogeneity factors (EHF) and control heterogeneity factors (CHF), with a total sample size of $N=500$ patients and significance level $\alpha = 0.025$. anova+t = anova contrast tests with $t$-approximation, anova+boot = anova tests with bootstrap approximation, marg+t = marginal tests with $t$-approximation, marg+boot = marginal tests with bootstrap, marg+shr+boot = marginal tests with shrinkage and bootstrap, strat+boot = stratified estimate with bootstrap, rem = random effects model.}
\label{tab:fwer}
\end{table}

As we have already seen in Example \ref{ex1}, the FWER of the ANOVA tests with the $t$- approximation from section \ref{sec:anova} turns out to be highly inflated, to an extent depending on the inhomogeneity of the subgroup effects. For instance, with EHF = 10, the FWER rises to 0.3145, far exceeding the nominal $\alpha$-level. In contrast, the bootstrap method applied to the ANOVA test statistics yields a considerably smaller FWER, but under high effect heterogeneities, the error is no longer controlled (FWER = 0.0337 for EHF = 10). The marginal tests with the $t$-approximation (using the Satterthwaite-formula) consistently control the FWER under the target level, regardless of effect or control heterogeneity, but they become very conservative under high heterogeneities. The bootstrap-approximation with the marginal tests performs similarly to the ANOVA bootstrap tests, but performs somewhat worse under high effect heterogeneity. Shrinkage applied to the marginal tests gives no difference compared to bootstrapping the ANOVA tests. The simulated FWER values of the stratified estimator are comparable to those of the marginal tests. Finally, the random effect models are extremely conservative throughout, primarily due to convergence issues. Overall, the bootstrap approximation for the ANOVA tests shows the best performance in these settings. 

For $N=250$ patients, correspondingly higher error probabilities are observed, but the comparison of the methods is similar. Detailed results can be found in Appendix \ref{app:res}. For $N=1000$, all bootstrap-based tests achieve values noticeably closer to the target $\alpha = 0.025$. In particular, the bootstraped ANOVA tests perform relatively well even under high effect heterogeneity, with an FWER of 0.0284 for EHF = 10. On the other hand, this still corresponds to a deviation from the target of more than 10\%. We did the same simulations also in the case with overlapping target populations (instead of nested populations) and found no particaular differences in the results. 

In further simulations we also applied a 2:1 randomization to the treatments in all three strata. Furthermore, we also applied a 2:1 randomization only in the intersection of the target populations (and a 1:1 allocation otherwise), to model the situation where two distinct treatments are tested. The corresponding results for $N=500$ patients are shown in Appendix \ref{app:res}. While they are very similar to the equal allocation cases under the 2:1 assigment in all strata, when restricting on the intersection only the stratified tests stay robust to increasing the control group heterogeneity. The other bootstrap methods thereby become more conservative, while the marginal tests using the t-approximation become excessively liberal. 

Another modification that we did was to assume only equal allocation probabilities to the treatments within the strata, instead of using an exactly equal allocation. This is achieved by drawing all subgroup-treatment specific sample sizes from a corresponding multinomial distribution, instead of just drawing the subgroup-specific sample sizes. In that case, for $N=500$, the different tests are all becoming more liberal, such that only the marginal tests with the $t$-approximation control the FWER across all settings. See the detailed results in Appendix \ref{app:res} (allocation pattern \enquote{D}).

\subsection{Simulated power} \label{sec:sim-power}

Table \ref{tab:pow} shows the resulting power values, which correspond to the mean number of rejections divided by the number of hypotheses, for $N=500$ patients. They reflect the already observed different extents of FWER-exploitation and -exceedance quite well. A notable observation is that with very large control group response heterogeneities (CHF = 10), the marginal tests with the Satterthwaite-approximation seem to lose power particularly strongly compared to the other methods. No power values are given for EHF = 0 since this implies the global null hypothesis to hold.

\begin{table}[ht]
\centering
\begin{tabular}{cc|ccccccc}
\toprule
 EHF & CHF & anova+t & anova+boot & marg+t & marg+boot & marg+shr+boot & strat+boot & rem  \\
\midrule
1  & 0   & 0.9737 & 0.9702 & 0.9638 & 0.9699 & 0.9702 & 0.9698 & 0.1200 \\
1  & 1   & 0.9737 & 0.9702 & 0.9322 & 0.9694 & 0.9702 & 0.9698 & 0.0890 \\
1  & 10  & 0.9737 & 0.9702 & 0.0598 & 0.9639 & 0.9702 & 0.9698 & 0.0887 \\
10 & 0   & 0.9984 & 0.9934 & 0.9887 & 0.9940 & 0.9934 & 0.9940 & 0.0835 \\
10 & 1   & 0.9984 & 0.9934 & 0.9884 & 0.9940 & 0.9934 & 0.9940 & 0.0825 \\
10 & 10  & 0.9984 & 0.9934 & 0.9581 & 0.9936 & 0.9934 & 0.9940 & 0.0823 \\
\bottomrule
\end{tabular}

\vspace{10pt}
\caption{Simulated multiple power for different effect heterogeneity factors (EHF) and control heterogeneity factors (CHF). $N=500$, $\alpha = 0.025$. Abbreviations as in Table \ref{tab:fwer}.}
\label{tab:pow}
\end{table}

\section{Simultaneous confidence intervals} \label{sec:ci}

We can use the critical values and standard errors of the different tests from Section \ref{sec:tests} to derive corresponding simultaneous confidence intervals. For each $I \in \mathcal{I}$, let $c_I$ denote the critical value associated with the test. Depending on the approximation method, $c_I$ is either obtained from the Satterthwaite approximation in equation~(\ref{eq4}) or, in the case of bootstrap methods, as the $(1-\alpha)$-quantile of the empirical distribution of the bootstrapped $\max_{I \in \mathcal{I}} Z_I^*$. Moreover, let $\text{SE}_I$ be the standard error of the effect estimate $\hat{\theta}_I$ used. So for example, we have \[\text{SE}_I = \frac{1}{n_{\Pop_I}}\sqrt{\sigma^2 \sum_{i \in I} (\delta_{\S_i, E_I}^{-1} + \delta_{\S_i, C}^{-1})n_{\S_i} + \hat{\boldsymbol{\theta}}_{\S,I} \hat{\boldsymbol{\Sigma}} \hat{\boldsymbol{\theta}}_{\S,I}^T}\] for the stratified tests. The simultaneous confidence intervals for the unknown effects $\btheta = (\theta_I)_{I \in \I}$ are then given by: 
\[\left[\hat{\theta}_I - c_I \text{SE}_I, \infty\right), \quad I \in \I\] and they are FWER-adjusted in the situations where this is the case for the corresponding tests (see Section \ref{sec:sim}).

\section{Real data example} \label{sec:ex}

\begin{figure}[ht]
\centering
\includegraphics[width=0.85\textwidth]{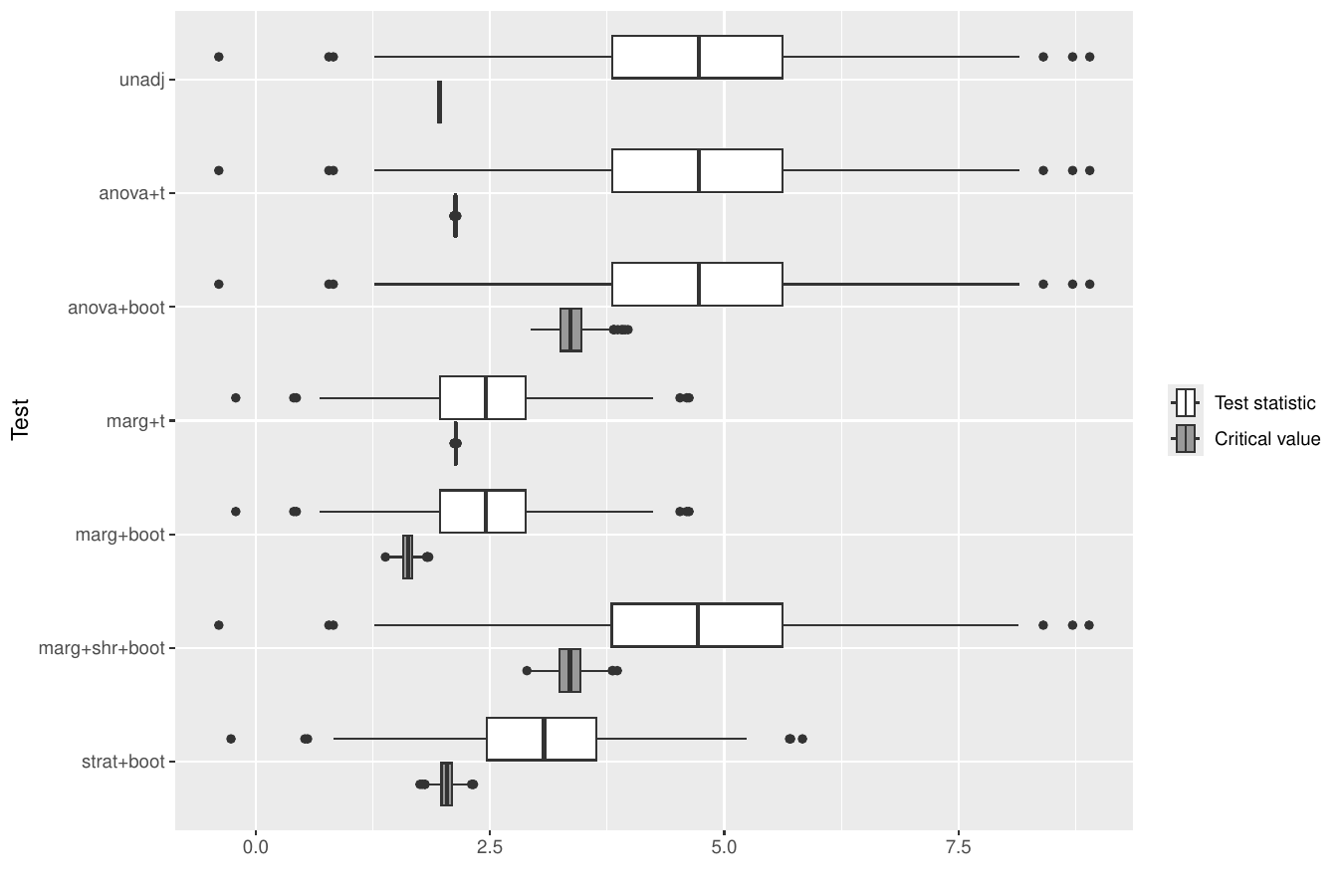}
\caption{Test statistics and critical values for unadjusted testing (unadj), the standard ANOVA tests (anova+t), the bootstrap method (anova+boot), the marginal tests with the t- (marg+t) and bootstrap approximation (marg+boot) and the stratified tests with the bootstrap approximation (strat+boot) in the real data example.}
\label{fig:boxplots}
\end{figure}

We compare the bootstrapped, marginal and stratified tests to the standard ANOVA tests and unadjusted testing in a real data example using a data set created by \citet{kesselmeier} from the MAXSEP study (\citet{brunkhorst}), which evaluated the effects of meropenem compared to a combination therapy of moxifloxacin and meropenem in patients with severe sepsis. The data consists of 1000 resamples of $N=500$ patients who were tested for two binary biomarkers $B_1$ (baseline lactate value $> 2 \text{ mmol}/\text{L}$) and $B_2$ (baseline C-reactive proteine value $> 128 \text{ mg}/\text{L}$), and who were randomly assigned to two arms of an umbrella trial. Assuming a positive effect of one unit in patients with a positive $B_1$ and $B_2$, and 0.25 points negative effects in the other strata, we computed the test statistics and critical values of the different tests in the overall population. 
The results are shown in Figure \ref{fig:boxplots}. One can see that the null hypothesis would often be rejected in unadjusted testing and in the standard ANOVA tests, whereas the bootstrap-method, the marginal tests, the shrinkage method and the stratified tests are noticeably more conservative. This aligns with the already observed degrees of FWER exhaustion in Section \ref{sec:sim-fwer}.

\section{Discussion} \label{sec:dis}

Treatment effects usually vary between patients, sometimes substantially. We have seen that a qualitative effect heterogeneity, where the effects differ in direction across subgroups, can compromise type I error control and necessitates a more conservative approach than doing the standard ANOVA contrast tests. We were able to reach this through various methods presented in Section \ref{sec:tests}. The best performance -- with the smallest exceedance of the significance level in different settings -- is achieved by the bootstrap approximation for the distribution of the ANOVA test statistics. However, it should be noted that noticeable exceedance of the significance level was still observed for the sample sizes examined, which decreases as the sample sizes increase. The marginal tests with the $t$-approximation may become overly conservative under high subgroup heterogeneities. The mixed effect models perform very poorly in general, probably due to misspecification of the random strata-wise effects, which are assumed to be redrawn from the normal distribution in every repetition of a study. It seems more natural instead to define them via a discrete random variable as we did for the marginal tests.

If the variance in a population is the same under the investigational and the control treatment, as is the case, for example, with non-predictive biomarkers (which do not provide any indication of how likely patients are to respond to the treatments), the marginal tests reduce to the $t$-test and the standard approach remains valid. The standard ANOVA tests also remain valid when testing the stronger intersection hypotheses $H_I = \cap_{i \in I}H_{\S_i}$, where $H_{\S_i} \colon \theta_{\S_i} \leq 0$ is the null hypothesis restricted to subpopulation $\S_i$. However, this reduces the validity of our testing in another sense, as a significant test result only indicates a treatment effect being present in at least one stratum, without identifying which one. Doing separate tests across all subpopulations may also be very ineffective, as this may greatly increase the number of hypotheses to be tested, and since the sample sizes within each stratum are typically much smaller than in the target populations. This can lead to a significant loss of power. 

Low power is generally a problem in multipopulation studies, especially when some of the subpopulations have small prevalences or are highly stratified. \citet{brannath} proposed an alternative type I error rate for such situations, the population-wise error rate (PWER). It is more liberal than the FWER while still controlling an average of the FWER restricted to the subpopulations. Therefore the issue discussed here also applies to PWER-control and can be handled analogously. The same holds for closed testing procedures and step-up or step-down methods.

We further note that the lack of type I error control in the ANOVA model also occurs in the special case of a single target population composed of heterogeneous subpopulations. This shows that the problem addressed here is not due to the multiplicity correction via FWER or PWER control, but rather to the bias in estimating the treatment effect in a target population when conditioning on its subgroup sample sizes.

Finally, we want to adress an apparent paradox which comes when defining treatment effects via a discrete random variable (instead of a constant) as we did in Section \ref{sec:marg}. If first a subpopulation realizes with a certain probability and then the patient outcomes are generated in this population, our observations would no longer be independent. This means that we would probably perform worse at FWER control than if we had known nothing about the heterogeneity of the strata. In fact, one should think differently: with each patient the pair consisiting of the response value and the strata membership is realized (in a single step) and therefore, we can still assume that the observations are independent. 

Our considerations may also be relevant in adaptive and group sequential study designs. We will adress this in our future research.

\section*{Acknowledgements}

We thank Miriam Kesselmeier for kindly providing the data set used in the real data example.

\bibliographystyle{unsrtnat}
\bibliography{references}

\appendix
\renewcommand{\thesubsection}{\Alph{subsection}}
\section*{Appendix}

\subsection{Convergence of the noncentrality parameter in Example \ref{ex1}} \label{app:ex1}

We regard $n_{\S_2}$ and $n_{\S_3}$ as a sum of i.i.d.\ Bernoulli random variables with parameters $1/6$ and $2/3$. Then by the CLT we have
\begin{align*}
\frac{\sqrt{N} \left(n_{\S_2}/N - 1/6 \right)}{\sqrt{1/6 \cdot 5/6}} \xrightarrow[N \to \infty]{d} N(0, 1) \quad &\Rightarrow \quad \frac{n_{\S_2}}{\sqrt{N}} \xrightarrow[N \to \infty]{d} N\left(\frac{1}{6}, \frac{5}{36}\right), \\
\frac{\sqrt{N} \left(n_{\S_3}/N - 2/3 \right)}{\sqrt{2/3 \cdot 1/3}} \xrightarrow[N \to \infty]{d} N(0, 1) \quad &\Rightarrow \quad \frac{n_{\S_3}}{\sqrt{N}} \xrightarrow[N \to \infty]{d} N\left(\frac{2}{3}, \frac{2}{9}\right)
\end{align*}

This implies that
\begin{align*}
\frac{4n_{\S_2} - n_{\S_3}}{\sqrt{N}} &\to N\left(0, 4^2 \cdot \frac{5}{36} + \frac{2}{9} + 2 \cdot 4 \cdot \underbrace{\operatorname{Cov}(n_{\S_2}, n_{\S_3})}_{1/6 \cdot 2/3}\right) = N\left(0,\frac{10}{3}\right)\\
\frac{4n_{\S_2} - n_{\S_3}}{\sqrt{5N/6}} &\to N\left(0, \frac{10}{3} \cdot \frac{6}{5}\right) = N(0,4).
\end{align*}
Also, we find that 
\begin{align*}
\sqrt{\frac{5N/6}{n_{\S_2} + n_{\S_3}}} = \sqrt{\frac{5/6}{(n_{\S_2} + n_{\S_3})/N}} \to \sqrt{\frac{5/6}{1/6 + 2/3}} =1
\end{align*}
by the law of large numbers.

\subsection{Variance of the stratified effect estimate} \label{app:strat}

Let $\boldsymbol{\theta}_{\S,I} = (\theta_{S_i, E_I})_{i \in I}$ denote the strata-wise effects, let $\delta_{\S_i, T}$ denote the allocation rate to the treatment $T$ in $\S_i$, and let $\boldsymbol{\Sigma}_I$ be the covariance matrix of the multinomial distribution $M(n_{\Pop_I}, \boldsymbol{\pi}_I)$ which is given by $\boldsymbol{\Sigma}_I = n_{\Pop_I} \cdot \left(\text{diag}(\boldsymbol{\pi}_I) - \boldsymbol{\pi}_I\boldsymbol{\pi}_I^T\right)$ for $\boldsymbol{\pi}_I = (\pi_{\S_i})_{i \in I}$. By the law of total variance, conditioning on the subgroup sample sizes $\boldsymbol{n}_I = \left(n_{\S_i}\right)_{i \in I}$, we find that

\begin{align*}
\Var\left(\sum_{i \in I} \frac{n_{S_i}}{n_{\Pop_I}} \left(\bar{Y}_{\S_i, E_I} - \bar{Y}_{\S_i, C}\right)\right) &= \E \left(\Var\left[\sum_{i \in I} \frac{n_{S_i}}{n_{\Pop_I}} \left(\bar{Y}_{\S_i, E_I} - \bar{Y}_{\S_i, C}\right) \middle| \boldsymbol{n}_I \right]\right) \\
&\quad  + \Var\left(\E\left[\sum_{i \in I} \frac{n_{S_i}}{n_{\Pop_I}} \left(\bar{Y}_{\S_i, E_I} - \bar{Y}_{\S_i, C}\right) \middle| \boldsymbol{n}_I \right]\right) \\
&= \E \left(\frac{\sigma^2}{n_{\Pop_I}^2} \sum_{i \in I} n_{\S_i}^2 \left(\frac{1}{n_{\S_i,E_I}} + \frac{1}{n_{\S_i,C}} \right)   \right) + \Var(\boldsymbol{n}^T \boldsymbol{\theta}_{\S,I})/n_{\Pop_I}^2 \\
&=\frac{\sigma^2}{n_{\Pop_I}^2} \sum_{i \in I} \left(\frac{1}{\delta_{\S_i, E_I}} + \frac{1}{\delta_{\S_i, C}}\right)n_{\Pop_I} \cdot \frac{\pi_{\S_i}}{\pi_{\Pop_I}} + \frac{\boldsymbol{\theta}_{\S,I} \boldsymbol{\Sigma}_I \boldsymbol{\theta}_{\S,I}^T}{n_{\Pop_I}^2},
\end{align*}
which is estimated by plugging in $n_{\S_i}$ for $n_{\Pop_I}  \pi_{\S_i}/\pi_{\Pop_I}$.

\clearpage

\subsection{Further simulation results} \label{app:res}

\subsection*{FWER}

\small

\begin{tabular}{cccc|ccccccc}
\toprule
$N$& alloc & EHF & CHF & anova & anova+boot & marg+t & marg+boot &  marg+shr+boot & strat+boot & rem \\
\midrule
250 & A & 0  & 0   & 0.0249 & 0.0265 & 0.0237 & 0.0273 & 0.0265 & 0.0269 & 0.0095 \\
250 & A & 0  & 1   & 0.0249 & 0.0265 & 0.0029 & 0.0260 & 0.0265 & 0.0269 & 0.0036 \\
250 & A & 0  & 10  & 0.0249 & 0.0265 & 0.0000 & 0.0254 & 0.0265 & 0.0269 & 0.0036 \\
250 & A & 1  & 0   & 0.0699 & 0.0325 & 0.0151 & 0.0327 & 0.0325 & 0.0339 & 0.0011 \\
250 & A & 1  & 1   & 0.0699 & 0.0325 & 0.0050 & 0.0325 & 0.0325 & 0.0339 & 0.0006 \\
250 & A & 1  & 10  & 0.0699 & 0.0325 & 0.0000 & 0.0334 & 0.0325 & 0.0339 & 0.0007 \\
250 & A & 10 & 0   & 0.3216 & 0.0446 & 0.0029 & 0.0479 & 0.0446 & 0.0487 & 0.0001 \\
250 & A & 10 & 1   & 0.3216 & 0.0446 & 0.0031 & 0.0492 & 0.0446 & 0.0487 & 0.0000 \\
250 & A & 10 & 10  & 0.3216 & 0.0446 & 0.0015 & 0.0486 & 0.0446 & 0.0487 & 0.0000 \\ \hline
1000 & A & 0  & 0   & 0.0255 & 0.0262 & 0.0251 & 0.0264 & 0.0262 & 0.0263 & 0.0096 \\
1000 & A & 0  & 1   & 0.0255 & 0.0262 & 0.0030 & 0.0261 & 0.0262 & 0.0263 & 0.0040 \\
1000 & A & 0  & 10  & 0.0255 & 0.0262 & 0.0000 & 0.0257 & 0.0262 & 0.0263 & 0.0040 \\
1000 & A & 1  & 0   & 0.0633 & 0.0281 & 0.0162 & 0.0285 & 0.0280 & 0.0289 & 0.0005 \\
1000 & A & 1  & 1   & 0.0633 & 0.0281 & 0.0053 & 0.0283 & 0.0281 & 0.0289 & 0.0003 \\
1000 & A & 1  & 10  & 0.0633 & 0.0281 & 0.0001 & 0.0286 & 0.0281 & 0.0289 & 0.0003 \\
1000 & A & 10 & 0   & 0.3106 & 0.0284 & 0.0036 & 0.0325 & 0.0284 & 0.0328 & 0.0000 \\
1000 & A & 10 & 1   & 0.3106 & 0.0284 & 0.0029 & 0.0325 & 0.0284 & 0.0328 & 0.0000 \\
1000 & A & 10 & 10  & 0.3106 & 0.0284 & 0.0021 & 0.0318 & 0.0284 & 0.0328 & 0.0000 \\ \hline
500 & B & 0  & 0   & 0.0255 & 0.0267 & 0.0242 & 0.0271 & 0.0267 & 0.0268 & 0.0097 \\
500 & B & 0  & 1   & 0.0255 & 0.0267 & 0.0030 & 0.0262 & 0.0267 & 0.0268 & 0.0039 \\
500 & B & 0  & 10  & 0.0255 & 0.0267 & 0.0000 & 0.0260 & 0.0267 & 0.0268 & 0.0039 \\
500 & B & 1  & 0   & 0.0622 & 0.0295 & 0.0123 & 0.0294 & 0.0295 & 0.0303 & 0.0010 \\
500 & B & 1  & 1   & 0.0622 & 0.0295 & 0.0054 & 0.0292 & 0.0295 & 0.0303 & 0.0005 \\
500 & B & 1  & 10  & 0.0622 & 0.0295 & 0.0000 & 0.0295 & 0.0295 & 0.0303 & 0.0006 \\
500 & B & 10 & 0   & 0.3073 & 0.0344 & 0.0017 & 0.0387 & 0.0344 & 0.0395 & 0.0001 \\
500 & B & 10 & 1   & 0.3073 & 0.0344 & 0.0012 & 0.0395 & 0.0344 & 0.0395 & 0.0000 \\
500 & B & 10 & 10  & 0.3073 & 0.0344 & 0.0025 & 0.0395 & 0.0344 & 0.0395 & 0.0000 \\\hline
500 & C & 0  & 0   & 0.0259 & 0.0285 & 0.0253 & 0.0290 & 0.0285 & 0.0272 & 0.0100 \\
500 & C & 0  & 1   & 0.6169 & 0.0270 & 0.3928 & 0.0272 & 0.0270 & 0.0272 & 0.0042 \\
500 & C & 0  & 10  & 0.9904 & 0.0074 & 0.7971 & 0.0127 & 0.0074 & 0.0272 & 0.0042 \\
500 & C & 1  & 0   & 0.2419 & 0.0275 & 0.1307 & 0.0288 & 0.0275 & 0.0301 & 0.0009 \\
500 & C & 1  & 1   & 0.5239 & 0.0265 & 0.3457 & 0.0277 & 0.0265 & 0.0301 & 0.0005 \\
500 & C & 1  & 10  & 0.9789 & 0.0154 & 0.7922 & 0.0174 & 0.0154 & 0.0301 & 0.0005 \\
500 & C & 10 & 0   & 0.4178 & 0.0378 & 0.2465 & 0.0373 & 0.0378 & 0.0394 & 0.0001 \\
500 & C & 10 & 1   & 0.4519 & 0.0356 & 0.2836 & 0.0362 & 0.0356 & 0.0394 & 0.0001 \\
500 & C & 10 & 10  & 0.8560 & 0.0327 & 0.5651 & 0.0293 & 0.0327 & 0.0394 & 0.0001 \\ \hline
500 & D & 0  & 0   & 0.0255 & 0.0265 & 0.0249 & 0.0269 & 0.0265 & 0.0267 & 0.0097 \\
500 & D & 0  & 1   & 0.1401 & 0.0268 & 0.0265 & 0.0265 & 0.0268 & 0.0267 & 0.0037 \\
500 & D & 0  & 10  & 0.5792 & 0.0165 & 0.0263 & 0.0198 & 0.0165 & 0.0267 & 0.0037 \\
500 & D & 1  & 0   & 0.0853 & 0.0294 & 0.0237 & 0.0299 & 0.0294 & 0.0304 & 0.0008 \\
500 & D & 1  & 1   & 0.1836 & 0.0295 & 0.0253 & 0.0296 & 0.0295 & 0.0304 & 0.0006 \\
500 & D & 1  & 10  & 0.5783 & 0.0235 & 0.0260 & 0.0255 & 0.0235 & 0.0304 & 0.0006 \\
500 & D & 10 & 0   & 0.3464 & 0.0360 & 0.0171 & 0.0389 & 0.0360 & 0.0401 & 0.0001 \\
500 & D & 10 & 1   & 0.3809 & 0.0357 & 0.0197 & 0.0393 & 0.0357 & 0.0401 & 0.0001 \\
500 & D & 10 & 10  & 0.5756 & 0.0360 & 0.0243 & 0.0383 & 0.0360 & 0.0401 & 0.0001 \\

\bottomrule
\end{tabular}

\clearpage

\normalsize
\subsection*{Power}

\small
\begin{tabular}{cccc|ccccccc}
\toprule
$N$& alloc& EHF & CHF & anova+t & anova+boot & marg+t & marg+boot & marg+shr+boot & strat+boot & rem \\
\midrule
250 & A & 1  & 0   & 0.9427 & 0.9350 & 0.9253 & 0.9343 & 0.9349 & 0.9339 & 0.1392 \\
250 & A & 1  & 1   & 0.9427 & 0.9350 & 0.8551 & 0.9342 & 0.9350 & 0.9339 & 0.0984 \\
250 & A & 1  & 10  & 0.9427 & 0.9350 & 0.0232 & 0.9183 & 0.9350 & 0.9339 & 0.0983 \\
250 & A & 10 & 0   & 0.9977 & 0.9891 & 0.9784 & 0.9898 & 0.9891 & 0.9898 & 0.0845 \\
250 & A & 10 & 1   & 0.9977 & 0.9891 & 0.9768 & 0.9900 & 0.9891 & 0.9898 & 0.0806 \\
250 & A & 10 & 10  & 0.9977 & 0.9891 & 0.9293 & 0.9867 & 0.9891 & 0.9898 & 0.0800 \\\hline
1000 & A & 1  & 0   & 0.9884 & 0.9872 & 0.9853 & 0.9871 & 0.9872 & 0.9871 & 0.1034 \\
1000 & A & 1  & 1   & 0.9884 & 0.9872 & 0.9674 & 0.9859 & 0.9872 & 0.9871 & 0.0810 \\
1000 & A & 1  & 10  & 0.9884 & 0.9872 & 0.2372 & 0.9823 & 0.9872 & 0.9871 & 0.0809 \\
1000 & A & 10 & 0   & 0.9993 & 0.9956 & 0.9924 & 0.9961 & 0.9956 & 0.9961 & 0.0836 \\
1000 & A & 10 & 1   & 0.9993 & 0.9956 & 0.9913 & 0.9961 & 0.9956 & 0.9961 & 0.0841 \\
1000 & A & 10 & 10  & 0.9993 & 0.9956 & 0.9800 & 0.9961 & 0.9956 & 0.9961 & 0.0842 \\\hline
500 & B & 1  & 0   & 0.9697 & 0.9663 & 0.9546 & 0.9653 & 0.9663 & 0.9656 & 0.1221 \\
500 & B & 1  & 1   & 0.9697 & 0.9663 & 0.9197 & 0.9666 & 0.9663 & 0.9656 & 0.0893 \\
500 & B & 1  & 10  & 0.9697 & 0.9663 & 0.0564 & 0.9598 & 0.9663 & 0.9656 & 0.0891 \\
500 & B & 10 & 0   & 0.9983 & 0.9932 & 0.9844 & 0.9939 & 0.9932 & 0.9939 & 0.0823 \\
500 & B & 10 & 1   & 0.9983 & 0.9932 & 0.9839 & 0.9940 & 0.9932 & 0.9939 & 0.0816 \\
500 & B & 10 & 10  & 0.9983 & 0.9932 & 0.9509 & 0.9939 & 0.9932 & 0.9939 & 0.0815 \\ \hline
500 & C & 1  & 0   & 0.9709 & 0.9688 & 0.9660 & 0.9691 & 0.9688 & 0.9672 & 0.1219 \\
500 & C & 1  & 1   & 0.9960 & 0.9512 & 0.9846 & 0.9472 & 0.9512 & 0.9672 & 0.0906 \\
500 & C & 1  & 10  & 1.0000 & 0.8668 & 0.9670 & 0.8942 & 0.8668 & 0.9672 & 0.0905 \\
500 & C & 10 & 0   & 0.9972 & 0.9912 & 0.9864 & 0.9928 & 0.9912 & 0.9939 & 0.0827 \\
500 & C & 10 & 1   & 0.9981 & 0.9910 & 0.9888 & 0.9925 & 0.9910 & 0.9939 & 0.0808 \\
500 & C & 10 & 10  & 1.0000 & 0.9784 & 0.9987 & 0.9785 & 0.9784 & 0.9939 & 0.0805 \\ \hline
500 & D & 1  & 0   & 0.9724 & 0.9698 & 0.9628 & 0.9691 & 0.9698 & 0.9697 & 0.1193 \\
500 & D & 1  & 1   & 0.9666 & 0.9586 & 0.9256 & 0.9641 & 0.9586 & 0.9697 & 0.0884 \\
500 & D & 1  & 10  & 0.8402 & 0.9001 & 0.2030 & 0.9130 & 0.9001 & 0.9697 & 0.0881 \\
500 & D & 10 & 0   & 0.9977 & 0.9933 & 0.9877 & 0.9939 & 0.9933 & 0.9941 & 0.0835 \\
500 & D & 10 & 1   & 0.9974 & 0.9932 & 0.9871 & 0.9938 & 0.9932 & 0.9941 & 0.0819 \\
500 & D & 10 & 10  & 0.9949 & 0.9888 & 0.9556 & 0.9910 & 0.9888 & 0.9941 & 0.0817 \\

\bottomrule
\end{tabular}

$N$ = sample size, alloc = allocation pattern (A = 1:1 in all strata, B = 2:1 in all strata, C = 2:1 in intersection, 1:1 in others, D = probabilistic 1:1 in all strata), EHF = effect heterogeneity factor, CHF = control heterogeneity factor, anova+t = anova contrast tests with $t$-approximation, anova+boot = anova tests with bootstrap approximation, marg+t = marginal tests with $t$-approximation, marg+boot = marginal tests with bootstrap, marg+shr+boot = marginal tests with shrinkage and bootstrap, strat+boot = stratified estimate with bootstrap, rem = random effects model

\end{document}